\documentclass[12pt]{article}

\usepackage{graphicx}
\usepackage{epsfig}
\newcommand{\be}{\begin{eqnarray}}
\newcommand{\ee}{\end{eqnarray}}

\begin{document}
%\titlerunning{Title running}
\begin{center}
{\Large\bf \boldmath  THERMAL EFFECTS FOR QUARK AND GLUON DISTRIBUTIONS IN 
HEAVY-ION COLLISIONS
} %<== title (bold face, capitalize)

\vspace*{6mm}
{G.I.Lykasov, A.N.Sissakian, A.S.Sorin, O.V.Teryaev  }\\      %<== authors
{\small \it  JINR, Dubna, 141980, Russia}      %<== institutions
%            $^b$ Institution2}
\end{center}

\vspace*{6mm}

% abstract
\begin{abstract}

In-medium effects for distributions of quarks and
gluons in central A+A collisions are considered.
We suggest a duality principle, which means similarity of thermal 
spectra of hadrons produced in heavy-ion collisions and inclusive 
spectra which can be obtained within the dynamic quantum scattering
theory. Within the suggested approach we show that the mean square of 
the transverse momentum for these partons grows and then saturates
when the initial energy increases.   
It leads to the energy dependence of hadron transverse mass
spectra which is similar to that observed in heavy ion collisions.

\end{abstract}

\vspace*{6mm}
\section{Introduction}

Searching for a new physics in heavy-ion collisions at AGS, SPS
and RHIC energies has led to intense theoretical and experimental
activities in this field of research \cite{QM}.
In this respect the search for signals of a possible transition of
hadrons into the QCD predicted phase of deconfined  quarks and gluons,
quark-gluon plasma (QGP), is of particular interest.
One of these signals can be the recent experimental observation of
the transverse-mass spectra of kaons and pions
from central Au+Au and Pb+Pb collisions which revealed "anomalous"
dependence on the incident energy. The inverse slope parameter of the 
transverse mass distribution (the so called effective transverse temperature) 
at the mid-rapidity rather fast increases with incident energy in the AGS 
domain~\cite{Ah00}, then saturates at the SPS \cite{NA49} and RHIC energies 
~\cite{RHIC}.

In this paper we would like to discuss the physical meaning of the so-called thermal spectra
of hadrons produced in heavy-ion collisions, see for example \cite{Br-Munz,Cleymans}, and try to understand
the dynamic reason of such inclusive spectra. Then we focus on a possible theoretical interpretation
of the nontrivial energy dependence for the inverse slope parameter of the transverse mass spectra
of mesons produced in central heavy-ion collisions. 

\section{Duality principle}

According to many experimental data, inclusive spectra of hadrons produced in heavy-ion collisions can be
fitted by the Fermi-Dirac distribution, corresponding to the thermodynamic equilibrium ({\bf LE}) for 
the system of final hadrons, see for example \cite{Br-Munz,Cleymans} 
\be
f_h^{A}~=~ C^{A}_T\left\{\exp((\epsilon_h-\mu_h)/T)~\pm~1\right\}^{-1},     
\label{def:LE}
\ee
where $+$ is for fermions and $-$ is for bosons, $\epsilon_h$ and $\mu_h$ are the kinetic energy and 
the chemical potential of the hadron $h$, $T$ is the temperature, $C^{A}_T$ is the normalization coefficient 
depending on $T$. Actually, the parameter $T$ depends on the incident energy $\sqrt{s}$ in the $N-N$ c.m.s.
For mesons simplifying this case we can assume that $\mu_h \simeq 0$, (in fact, it generally cannot be strictly 
zero \cite{Sinyukov:02}); then Eq.(\ref{def:LE}) is usually presented in the form
\be
f_h^{A}~\simeq~ C^{A}_T\exp(-\epsilon_h/T)~.     
\label{def:LEmes}
\ee
On the other hand, according to the Regge theory and the $1/N$ expansion in QCD,
the inclusive spectrum of  hadrons produced, for example in $N-N$ collisions  at high energies,
has the scaling form, e.g., it depends only on $M_X^2/s$, where $M_X$ is the missing mass of 
produced hadrons, $s$ is the initial energy squared in the $N-N$ c.m.s.,
 and $M_X^2/s=1-x_r$, where $x_r=2E^*_h/\sqrt{s}$ is the radial Feynman variable,
$E^*_h$ is the energy of the hadron $h$ in the $N-N$ c.m.s.
For example, the quantum scattering theory and the fit of the experimental data for inclusive meson 
spectra at low $x_r$ results in
\be
\rho^{NN}_m(x_r)\sim C_N(1-x_r)^{d_N}
\label{def:spxr}
\ee
If $x_r<<1$, Eq.(\ref{def:spxr}) can be presented in the exponential form 
%At low values of $x_r<1/d$ or $E_h<\sqrt{s}/2d$ 
\be
\rho_m^{NN}\sim C_N exp(-d_Nx_r)
\label{def:xrexp}
\ee
Inserting the form for $x_r$ in Eq.(\ref{def:xrexp}), we get the inclusive spectrum of mesons in the form
similar to that of the thermal spectrum given by Eq.(\ref{def:LEmes}) 
\be
\rho_m^{NN}~=~C_N\exp(-d_N x_r)\equiv C_N\exp(-\frac{E^*_h}{T^N_s})~,
\label{def:DANN}
\ee
where $T^N_s=\sqrt{s}/2d_N$.
However, in contrast to Eq.(\ref{def:LEmes}), the form of the inclusive spectrum of mesons produced in $N-N$
collisions given by Eq.(\ref{def:DANN}) does not assume introduction of any temperature of mesons like $T$. 
Figure 1 illustrates the approximate equivalence between $\rho_m^{NN}$ given by Eq.(\ref{def:spxr}) and 
$\rho_m^{NN}$ given by Eq.(\ref{def:DANN}).
One can see from Fig.1 that at high energies these two forms for the meson spectrum are very similar to each other
to the meson energies about a few GeV. Therefore, $\rho_m^{NN}$ can be presented in the exponential form at low
and even moderate energies $E^*_h$.

Let us assume that in central $A-A$ collisions at high energies in the first $N-N$
interaction at some time the mini-jet consisting of $(q{\bar q})$ pairs is created and then
pions are produced in the central rapidity region. We also suggest that the distribution of
these pairs in the $(q{\bar q})$ mini-jet has the form similar to the one given by  
 Eq.(\ref{def:DANN})   
\be
f_{q\bar q}^{jet}~=~C_A(1-x_r)^{d_A}\simeq C_A\exp(-\frac{E^*_h}{T^A_s})~,
\label{def:DAAA}
\ee   
where $T^A_s=\sqrt{s}/2d_A$. In the general case, the parameter $d_A$ is not the same as $d_N$ which can be found
from the quantum scattering theory or fitting the experimental data on inclusive spectra of mesons produced
in $N-N$ collisions. Let us call the assumption corresponding to Eq.(\ref{def:DAAA}) the 
{\bf Dynamic ansatz (DA)}. One can suggest the {\bf duality principle} which is the similarity of thermal spectra
given by Eq.(\ref{def:LEmes}) and the dynamical spectra given by Eq.(\ref{def:DAAA}).
To find the form for $d_A$ one can use the approach suggested by Kuti, Weiskopf \cite{KW} for the
calculation of parton distribution in a nucleon. One can show that $d_A$ is proportional to the 
number $n$ of $q{\bar q}$ pairs in the mini-jet, e.g., $d_A\sim n$. To estimate the in-medium effects 
we replace $n$ by the mean multiplicity $<n>_\pi^{NN}$ of pions produced in $NN$ collision,
e.g., $d_A\simeq d_0<n>_\pi^{NN}$. 
\vspace*{0.5cm}
\begin{figure}[t]
\rotatebox{270}%
{\epsfig{file=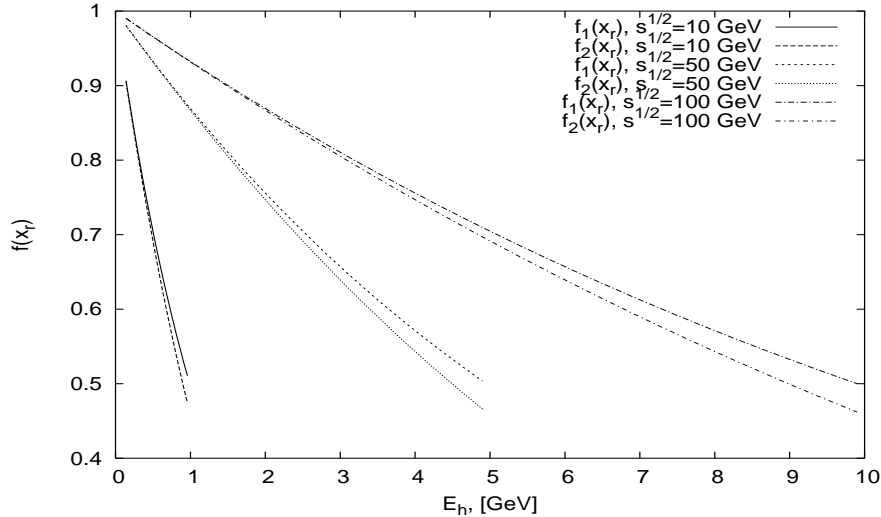,height=12cm,width=7cm }}
\caption[Fig.1]{$f_1(x_r)=exp(-dx_r)$ and  $f_2(x_r)=(1-x_r)^d$ as functions of $E_h$.}
\end{figure}
\section{Parton distribution in medium}

Recently the parton distribution in a medium was analyzed on the assumption of the local thermodynamic equilibrium 
for quark objects like hadrons produced in heavy ion central collisions \cite{lsst:08}. It was shown that, 
for example, the valence quark distribution in the quark object like the hadron $h$, which is 
in  local thermodynamic equilibrium with surrounding nuclear matter, can be calculated by the following 
equation:  
\be
f_{q_v}^{A}(x,{\bf p}_t)= \int_0^1dx_1\int_0^1dx_h\int d^2p_{1t}d^2p_{ht}
q_v^h(x,{\bf p}_t)q_r^h(x_1,{\bf p}_{1t})\times
\\ \nonumber
f_h^{A}(x_h,{\bf p}_{ht})\times 
\delta(x+x_1-x_h)\delta^{(2)}({\bf p}_t+{\bf p}_{1t}-{\bf p}_{ht})~,
\label{def:pdm}
\ee
where $f_h^{A}(x_h,{\bf p}_{ht})$ is the distribution of quark objects like hadrons locally equilibrated 
in a medium ({\bf LE}); $q_v^h, q_r^h$ are the probabilities to find the valance quark and other partons 
(valence, sea quarks (antiquarks) and gluons) in $h$; $x_1,x_h,x$ are the
Feynman variables, ${\bf p}_t,{\bf p}_{1t},{\bf p}_{ht}$ are the transverse momenta.    
The thermodynamic distribution like Eq.(\ref{def:LEmes}) was assumed in \cite{lsst:08} 
for $f_h^{A}(x_h,{\bf p}_{ht})$.  
The same form for $f_{q_v}^{A}$ can be obtained suggesting the dynamic distribution ({\bf DA}) for $f_h^{A}$ given by
Eq.(\ref{def:DAAA}) instead of Eq.(\ref{def:LEmes}).
Assuming the factorized form for $f_q^h(x,p_t)=f_q(x)g_q(p_t)$ we have approximately \cite{lsst:08} the following form
for the mean transverse momentum squared of the valence quark in a medium:
\be
<p_{q,t}^2(x\simeq 0)>_{q,appr.}^A&\simeq&\frac{<p_t^2>_q^h+{\tilde T}\sqrt{m_h^2+s/4}}
{1+{\tilde T}\sqrt{m_h^2+s \ / \ 4}/(2<p_t^2>_q^h)}~,
\label{def:avptsq}
\ee
where $\tilde T=T$ for Eq.(\ref{def:LEmes}) ({\bf LE}) and $\tilde T=T_s=\sqrt{s}/2d_A$ for {Eq.(\ref{def:DAAA}) 
(\bf DA}).
As is seen from Eq.(\ref{def:avptsq}) $<p_{q,t}^2>_{q,appr.}^A$ grows when $\sqrt{s}$ increases and then 
saturates, its more careful calculation is presented in \cite{lsst:08}. Note that in the {\bf LE} case 
$\sqrt{s}$ is some scale energy which cannot be equal to the initial energy $\sqrt{s_{NN}}$ \cite{lsst:08}, 
whereas in the {\bf DA} case it is the same as $\sqrt{s_{NN}}$. 
For mesons produced in central $A-A$ collisions we have similar broadening for the hadron $p_t$-spectrum
\cite{lsst:08}
\be
<p_{{h_1}t}^2>_{appr.}^{AA}\simeq
\frac{<p_{h_1t}^2>^{NN}/(1+r)+{\tilde T}\sqrt{m_m^2+s/4}}{1+{\tilde T}\sqrt{(m_m^2+s/4)}(1+r)\ / \
(2<p_{h_1t}^2>^{NN})}+ \frac{<p_{h_1t}^2>^{NN}}{r}~, 
\label{def:avptsh}
\ee
where $<p_{{h_1}t}^2>$ is the mean value for the transverse momentum squared of the meson $h_1$ produced in 
the central heavy-ion collision, $<p_t^2>_q^m$ is the same quantity for a quark in a medium, 
$r=\gamma_c/\gamma_q$, ${\tilde T}=T$ ({\bf  LE}) or ${\tilde T}=\sqrt{s}/2d_A$. Here $\gamma_q$ and 
$\gamma_c$ are the slopes in the Gauss form of the $p_t$ dependence for the quark distribution in the hadron 
$h$ and its fragmentation function, see details in \cite{lsst:08}.
As is seen from Eqs.(\ref{def:avptsq},\ref{def:avptsh}), the saturation properties for $<p_{q,t}^2>_q^A$
and $<p_{{h_1}t}^2>^{AA}$ at high $\sqrt{s}$ do not depend on the
 values of $\tilde T$, whereas the growth of these quantities at $\sqrt{s}\leq 20-30$ GeV is very sensitive 
to the value of $d_A$. 
To describe the experimental data on the transverse momentum squared of $K$-mesons
produced in central $A-A$ collisions we took $d_0=0.5$ and the energy dependence for $<n>_\pi^{NN}$ from
\cite{meannpi:1993}. We included also the energy dependence for the mean squared of the 
transverse momentum of kaons produced in the $N-N$ collision, see \cite{Br-Munz:2006} and references 
there in. 
 
In Fig.2 we present our estimation for the mean transverse momentum squared of the $K^+$-meson produced 
in the A-A collision. One can see from Fig.2 tat this quantity increases when the incident energy increases
to the AGS energies and then saturates at higher energies. Note that this calculation is
very approximate and we need to improve it including standard nuclear effects like rescattering
and others. The suggested approach results in the saturation of the effective slope $T_{eff}$ for the 
transverse mass spectrum of mesons produced in central heavy-ion collisions that is directly related 
to the quantity presented in Fig.2. In contrast to this the thermodynamic models predict the increase 
in $T_{eff}$ when $\sqrt{s_{NN}}$ increases even to very high energies \cite{Sinyukov}. Therefore, the
presented results can be verified by more careful measurements at the SPS energy and future 
experiments at the LHC.      
 
\begin{figure}[t]
\rotatebox{270}%
{\epsfig{file=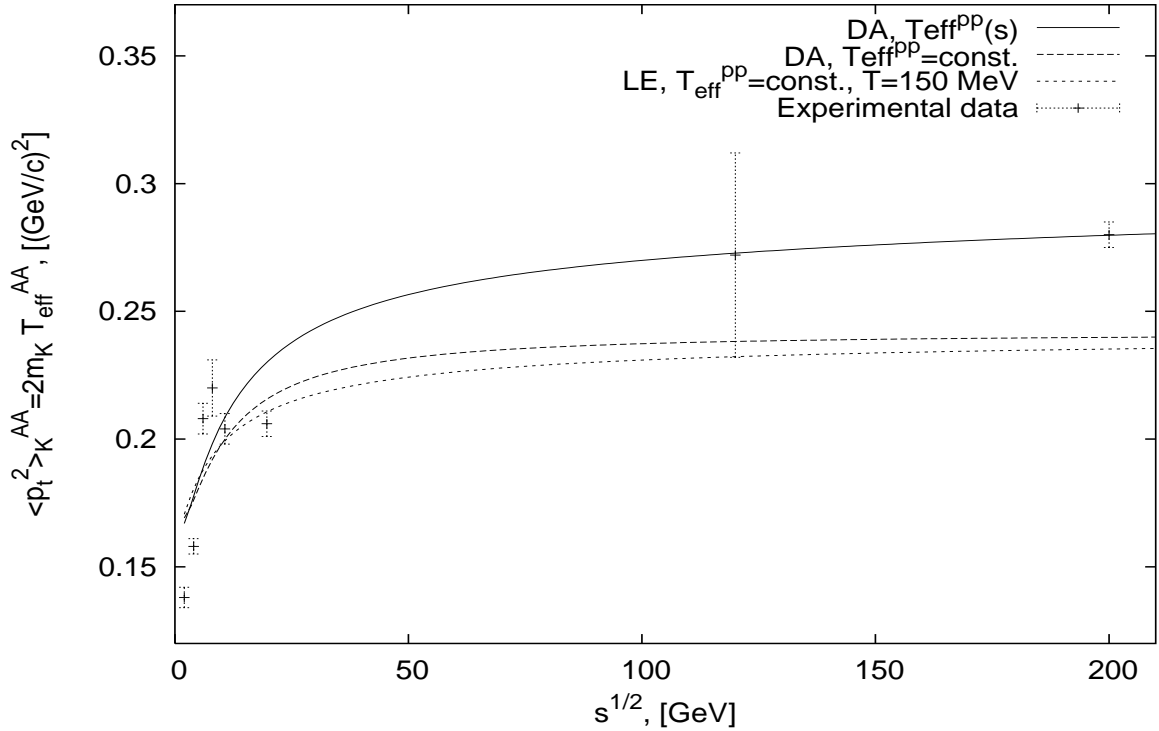,height=16.cm,width=10.cm }}
\caption[Fig.4]{The mean transverse momentum squared of the $K^+$-meson produced 
in the central A-A collision. The solid line corresponds to the DA, when the effective
slope for the transverse mass spectrum of kaons produced in the $p-p$ collision
$T_{eff}^{pp}$ depends on $\sqrt{s}$ \cite{Br-Munz:2006}. The long dashed line corresponds
to the DA, when $T_{eff}^{pp}=constant$, whereas the short dashed line corresponds
to the LE, when $T_{eff}^{pp}=constant$ and the temperature $T=150$ MeV. Experimental data
were taken from \cite{Ah00,NA49,RHIC}.}
\end{figure}

\section{Conclusion}

We suggest the duality principle. 
Thermal spectra of hadrons produced in central A-A collisions can have a
dynamical nature. Similar spectra can be obtained within the quantum scattering theory without
introducing the temperature. One can assume that hadron jets consisting of colorless quark 
objects are produced in central A-A collisions.
Then we get broadening for the mean transverse momentum squared 
of quarks in a medium respective to the incident energy.
Similar effects can be obtained for transverse momentum spectra of mesons produced
in central $A-A$ collisions.
The mean transverse momentum squared of these mesons as a function of the incident energy
grows and then saturates at high energies.

\vspace{0.25cm}
{\bf Acknowledgments}\\
The authors are grateful to A.Andronic, P.Braun-Munzinger, A.V.Efremov, L.L.Frankfurt,  
M.Gazdzicki, S.B.Gerasimov, J.Cleymans, A.B.Kaidalov, H.Satz, Yu.Sinyukov and V.D.Toneev  
for very useful discussions. This work was supported in part by RFBR 
project No 08-02-01003 and by the special program of the Ministry of Education and 
Science of the Russian Federation (grant RNP.2.1.1.5409).

\end{document}